\documentclass[11pt]{article}

\usepackage[T1]{fontenc}
\usepackage[utf8]{inputenc}
\usepackage[margin=1in]{geometry}
\usepackage{cite}
\usepackage{amsmath,amssymb,amsfonts}
\usepackage[ruled,vlined,linesnumbered]{algorithm2e}
\usepackage{graphicx}
\usepackage{textcomp}
\usepackage{xcolor}
\usepackage{booktabs}
\usepackage[most]{tcolorbox}
\usepackage{url}
\usepackage[hidelinks]{hyperref}

\providecommand{\keywords}[1]{%
  \par\vspace{0.75em}
  \noindent\textbf{Keywords:} #1
}

\begin{document}

\title{Benchmarking System One Models in Online Moderation}

\author{
Federico Mazzoni\\
Independent Researcher\\
\texttt{mazzoni.federico1@gmail.com}
\and
Andrea Failla\\
Institute of Information Science and Technologies (ISTI),\\
National Research Council (CNR), Italy\\
\texttt{andrea.failla@isti.cnr.it}
}

\date{}

\maketitle

\begin{abstract}
\noindent Online moderation systems must apply changing platform policies, community rules, and prior decisions while producing decisions that can be audited and routed to human review. We evaluate whether System One Models, which accept natural-language context but return typed choices, probabilities, or scores, can support this setting. Across five moderation benchmarks, \textit{Jev} is competitive with specialized or published reference systems, matching or exceeding them in several policy-grounded and harmful-content settings. We then use controlled information conditions to separate written rules, retrieved precedents, and restrictions on the candidate answer space. Jev generally benefits from retrieved precedents, improving exact policy selection and harmful-content discrimination when the answer space is held fixed. \textit{Laya} is less consistent: retrieval often shifts its positive prediction rate or no-violation rate without improving discrimination. Jev confidence can support selective review in several settings, although it is not consistently calibrated; Laya confidence is less useful for ranking errors. These results show a promising future for SOM-powered content moderation.
\end{abstract}

\keywords{System One Models; content moderation; retrieval-augmented decisions; Jev; Laya}

\section{Introduction}

Online platforms must continuously decide which user-generated content can remain visible, which content requires intervention, and which policy justifies that intervention. 
At the scale of contemporary social media, these decisions cannot rely exclusively on manual review, motivating extensive research on automated content moderation and harmful-content detection~\cite{gorwa2020algorithmic,schmidt2017survey,fortuna2018survey,contentmoderationreview2022}. 
Automated systems have consequently been developed for tasks ranging from offensive-language and hate-speech detection to the enforcement of broader platform policies. 
Yet content moderation is not simply a classification problem. What constitutes a violation depends on the policy being enforced, the community in which the content appears, and the contextual interpretation of both the message and the governing rules~\cite{gorwa2020algorithmic,fortuna2018survey}. 
The same expression may therefore receive different moderation decisions under different normative frameworks.
\\ \ \\
Traditional Machine Learning (ML) approaches provide an effective mechanism for detecting predefined categories of harmful content, and specialized classifiers remain strong baselines for tasks such as toxicity and offensive-language detection~\cite{schmidt2017survey,fortuna2018survey,hanu2020detoxify}. 
Their decision boundaries, however, are generally learned with respect to a fixed annotation scheme and training distribution. 
Changes in moderation policy, community norms, or violation taxonomy can consequently require new annotated data or model adaptation. 
Recent Large Language Models (LLMs) provide a different possibility: rather than encoding the moderation standard only during training, they can receive natural-language policies and examples directly at inference time. 
This has motivated emerging approaches to \emph{policy-as-prompt} moderation and, more broadly, growing interest in LLM-based moderation systems~\cite{palla2025policy,chen2025llmmoderation}. 
Such systems make it possible to condition decisions on changing policies and demonstrations, although their suitability for operational moderation also depends on whether their outputs can be made predictable, interpretable, and useful for downstream decision pipelines.
\\ \ \\
System One Models (SOMs) offer an interesting additional design point in this space. 
SOMs accept unstructured textual states and task descriptions, similarly to language models, but map them directly to typed, non-autoregressive outputs such as categorical choices, probabilities, or ordinal scores. 
Jev and Laya, the two SOM implementations considered in this work, therefore combine an interface capable of consuming natural-language context with structured outputs resembling those of conventional classifiers~\cite{typesafe2026jev,laya2026}. 
For moderation, this combination is particularly appealing: a system could potentially interpret a written policy or previously moderated examples while still returning an explicit decision and associated probability rather than unrestricted generated text.
\\ \ \\
Our contribution is threefold. First, we provide a controlled framework for evaluating content moderation under distinct sources of decision context, separating written policy, retrieved precedents, and restrictions on the available answer space. 
Second, we provide, to our knowledge, the first broad evaluation of SOMs for online content moderation, covering both exact policy enforcement and conventional harmful-content detection across five benchmarks. 
Third, we evaluate whether the probabilities and confidence values returned by typed SOM interfaces are sufficiently informative to support selective moderation and human review.
Our experiments address four research questions. \emph{RQ1} asks whether SOMs can apply written moderation policies and previously decided cases. \emph{RQ2} asks whether retrieved precedents improve the underlying decision signal or primarily alter the operating point or available answer space. \emph{RQ3} asks whether SOM probabilities and confidence scores provide useful estimates of uncertainty for moderation workflows. Finally, \emph{RQ4} asks whether the observed behavior extends from explicit policy enforcement to generic harmful-content detection.
The rest of the paper is organized as follows. Section~\ref{sec:related} reviews automated and policy-grounded content moderation, retrieval-based decision support, and System One Models. Section~\ref{sec:model} introduces the benchmark tasks, datasets, SOM information conditions, and output interfaces. Section~\ref{sec:eval} presents the experimental results and uncertainty analyses. Finally, Section~\ref{sec:conclusion} discusses the implications, limitations, and directions for future work.

\section{Related Work}
\label{sec:related}
In the following, we review relevant work in content moderation, in-context learning and SOMs.

\noindent\textbf{Content moderation.}
Automated moderation research has often been organized around detecting predefined categories of harmful text. Benchmarks such as OLID and HateXplain fix annotation schemes for offensive language, hate speech, and related labels~\cite{zampieri2019olid,mathew2021hatexplain}, while specialized systems such as Detoxify expose toxicity-oriented category scores or binary decisions~\cite{hanu2020detoxify}. Industrial moderation systems follow a similar fixed-taxonomy template at larger scale: the OpenAI moderation model, for example, is described as a classifier for categories such as sexual content, hateful content, violence, self-harm, and harassment, built with explicit taxonomies, labeling instructions, quality control, robustness checks, and active learning~\cite{markov2023holistic}.
\\
This paradigm has well-documented limits. Broad surveys emphasize that datasets differ in how they define hate, abuse, offense, and toxicity, which targets they cover, how labels are annotated, and which languages are represented~\cite{schmidt2017survey,fortuna2018survey,vidgen2020directions,poletto2021resources}. Even within a single dataset, separating hate speech from merely offensive language is difficult~\cite{davidson2017automated}. Reported performance can be inflated by dataset-specific artifacts and weak validation protocols~\cite{arango2019hate}, while functional tests expose failures that aggregate scores conceal, including reliance on identity terms and difficulty with counter-speech~\cite{rottger2021hatecheck}. Toxicity and hate-speech classifiers can also inherit annotation biases, producing disproportionate scores for identity-related or dialectal language~\cite{dixon2018measuring,sap2019risk,nogara2025toxic}.
\\
Platform-governance research offers a different diagnosis of this heterogeneity. Moderation is an institutional practice that defines acceptable participation and allocates work between automated systems and human reviewers~\cite{gillespie2018custodians,roberts2019behind,gorwa2020algorithmic}. At platform scale, moderation is better understood as a system of mass speech administration than as a sequence of isolated classification decisions~\cite{gillespie2020scale,douek2022systems}. The rules themselves vary substantially, as platforms publish different guidelines~\cite{jhaver2024communityguidelines} and e.g. Reddit communities maintain their own rule sets~\cite{fiesler2018reddit}. A single learned decision boundary is therefore an incomplete target when the operative standard changes across platforms, communities, and policies.
\\ \ \\
\noindent\textbf{Retrieval and in-context learning.}
Policy-conditioned moderation has also been studied as an interface problem. Binary question answering decomposes evolving moderation policies into explicit checks, making policy changes easier to localize than in a monolithic classifier~\cite{mullick2023evolving}. Llama Guard places a safety taxonomy directly in the prompt and uses an LLM-based safeguard model to classify unsafe inputs and outputs~\cite{inan2023llamaguard}. Policy-as-prompt work formalizes a related paradigm in which natural-language policies are supplied at inference time, while also highlighting sensitivity to prompt formulation, organizational governance, and accountability~\cite{palla2025policy}. Empirical studies of LLMs on community moderation report useful but uneven performance across rules and communities~\cite{kumar2024watch,kolla2024llmmod}, and reviews of LLM-based moderation describe prompting and fine-tuning as the main routes beyond fixed classifiers~\cite{chen2025llmmoderation}.
\\
Written rules are rarely sufficient by themselves; institutions often interpret them through prior decisions. Legal theory treats precedent as a commitment to decide like cases alike, while recognizing that the difficult judgment lies in deciding which features make two cases relevantly similar~\cite{schauer1987precedent}. Case-based reasoning formalizes related ideas computationally, including in legal argumentation~\cite{ashley1990modeling,kolodner1993case}. Moderation reseaarch has begun to use precedent in analogous ways. On Reddit, Crossmod used cross-community moderation decisions as signals to assist moderators in communities with limited local training data~\cite{chandrasekharan2019crossmod}.
\\
In current language-model systems, precedents enter through retrieval and in-context demonstrations. Retrieval-augmented generation (RAG) introduced the use of an external non-parametric memory to condition model outputs~\cite{lewis2020rag}. In-context learning research shows that what demonstrations communicate is not straightforward. Selection matters: semantically similar examples can outperform random ones~\cite{liu2022examples}, and learned retrievers can further improve demonstration selection~\cite{rubin2022learning}. Yet demonstrations can remain useful even when their input--label mappings are partly corrupted, because they also communicate format, input distribution, and label space~\cite{min2022demonstrations}. Related work distinguishes task recognition, in which examples identify a task the model already knows, from task learning, in which the model acquires the input--label mapping from the demonstrations themselves~\cite{pan2023what}; larger models are more capable of overriding semantic priors when labels are flipped or replaced by semantically unrelated tokens~\cite{wei2023larger}. Demonstrations can also induce majority-label, recency, and common-token biases~\cite{zhao2021calibrate}, and model performance is sensitive to demonstration order~\cite{lu2022fantastically}. 
\\ \ \\
\noindent\textbf{System One Models.}
System One Models are a recent advancement that consume natural-language states and task descriptions but return typed, non-autoregressive outputs, such as a choice over candidates, a probability for a proposition, or a score on an ordinal scale~\cite{typesafe2026jev,laya2026}. Their name echoes the fast, non-deliberative mode of judgment in dual-process accounts of cognition~\cite{kahneman2011thinking}. For moderation pipelines, this interface offers an interesting opportunity because it combines properties that usually belong to different system classes: like policy-as-prompt LLMs, SOMs can receive rules and precedents at inference time; like conventional classifiers, they return structured decisions and numerical scores directly, without parsing generated explanations.

\section{Methodology}
\label{sec:model}

We evaluate Jev v1.13 and Laya v0.3 on two related but distinct content-moderation tasks: \emph{policy-violation detection}, in which decisions are grounded in an explicit set of community or platform rules, and \emph{harmful-content detection}, in which the model directly predicts whether a piece of content belongs to a harmful or offensive class. We first formalize the two tasks and their associated benchmarks, and then introduce the information conditions and typed SOM interfaces used throughout the experiments.
Code and prompts used in this experiments are available in a dedicated GitHub repository~\footnote{\textbf{https://github.com/FedericoMz/som-moderation-benchmark}}.

\subsection{Benchmark Tasks and Datasets}

\noindent\textbf{Policy-violation detection.}
Let $x$ denote a target social-media message and let
$\mathcal{R}=\{r_1,\ldots,r_m\}$ denote the set of rules applicable to the target. The model predicts an outcome
\[
y \in \{0,1,\ldots,m\},
\]
where $y=0$ denotes \emph{No rules broken} and $y=j$ denotes a violation of rule $r_j$. This formulation induces two evaluation tasks. The first is binary \emph{violation detection},
\[
v = \mathbb{I}[y\neq 0],
\]
which asks whether any rule has been violated. The second is \emph{exact policy selection}, which additionally requires the model to identify the annotated rule when $v=1$.

We evaluate this task on two policy-grounded benchmarks:

\begin{itemize}
    \item \textbf{PluRule}~\cite{plurule2026} evaluates community-specific moderation on Reddit. Each instance contains two comments from the same discussion: one violates a named subreddit rule, while the other is treated as compliant with that rule. The dataset retains the rules and community metadata required to evaluate both binary violation detection and exact rule selection. We consider both a high-support setting based on the six largest hydrated communities and the published test-set setting across the complete set of available communities.

    \item \textbf{HateModerate}~\cite{zheng2024hatemoderate} evaluates enforcement of Facebook's hate-speech policy. Its examples are organized around 41 policy labels, allowing us to evaluate both whether a sentence violates the policy and which specific policy category applies. After grouping exact duplicates within each released split, removing two sentences with contradictory annotations, and excluding 175 sentences associated with multiple positive policies, which cannot be represented by our single-choice formulation, we retain 3,309 training sentences and 3,302 test sentences.
\end{itemize}

\medskip
\noindent\textbf{Harmful-content detection.}
The second task removes the explicit policy from the decision problem. Given a social-media item $x$, the model predicts
\[
h \in \{0,1\},
\]
where $h=1$ denotes harmful or offensive content according to the benchmark annotation. Unlike policy-violation detection, the model is not required to map the content to a rule. The task therefore measures whether the SOM can recover the benchmark's content-level decision boundary directly from the target and, when retrieval is enabled, from previously decided examples.

We consider three datasets:

\begin{itemize}
    \item \textbf{OLID}~\cite{zampieri2019olid} contains English tweets annotated for offensive-language detection. We use its first annotation level, distinguishing offensive (\texttt{OFF}) from non-offensive (\texttt{NOT}) content. After removing exact duplicates and contradictory annotations, the evaluation population contains 14,046 posts.

    \item \textbf{HateXplain}~\cite{mathew2021hatexplain} contains social-media posts annotated as \emph{hateful}, \emph{offensive}, or \emph{normal}. We restrict the analysis to the offensive and normal classes, yielding 14,123 posts after duplicate and contradiction removal.

    \item \textbf{ModerationBench}~\cite{majumdar2026moderationbench} contains real-world multimodal Bluesky posts with binary \emph{Safe}/\emph{Unsafe} moderation decisions. Because the evaluated SOMs operate on textual states, we construct a text-based adaptation by combining each post with a neutral description of its associated media generated by \texttt{OpenGVLab/InternVL3\_5-4B-Instruct}. Starting from 3,712 posts, we remove entries without usable textual or media content, contradictory exact-input groups, and redundant consistent duplicates. The resulting population contains 3,689 posts, of which 2,584 are labeled Safe and 1,105 Unsafe. Appendix~\ref{app:moderationbench-captions} reports sensitivity analyses under alternative caption generators.
\end{itemize}
 
For OLID and HateXplain, we additionally use the Detoxify family~\cite{hanu2020detoxify} as a task-specific external reference. Detoxify is a family of models trained on toxicity detection. Note that none of the datasets employed here was present in Detoxify's training data.

\subsection{SOM Information Settings}

We vary two sources of decision context: written \emph{normative} material and retrieved \emph{precedential} cases. Table~\ref{tab:information-settings} summarizes the resulting conditions and candidate-answer inventories.

\begin{table}[t]
\centering
\scriptsize
\setlength{\tabcolsep}{1.8mm}
\caption{Information settings used in the SOM evaluations. Rule-inventory variants apply only to policy-selection experiments.}
\label{tab:information-settings}
\begin{tabular}{lllp{5.5cm}}
\toprule
Condition & Written policy/rules & Retrieved precedents & Candidate answers \\
\midrule
Direct & No & No & Binary harmful-content labels \\
NAD & Yes & No & All rules or represented rules plus \emph{No rules broken} \\
RAD & No & Yes & Binary harmful-content labels, or retrieved violating rules plus \emph{No rules broken} \\
General & Yes & Yes & All, represented, or retrieved rules plus \emph{No rules broken} \\
\bottomrule
\end{tabular}
\end{table}

Retrieved precedents are selected by embedding messages, using as embedding model \texttt{qwen3-embedding:4b}~\cite{zhang2025qwenembedding}, and ranking admissible cases by cosine similarity. Unless otherwise specified, each target receives the five nearest precedents, excluding the target itself and respecting the task domain; for example, PluRule precedents must come from the same subreddit. For OLID and HateXplain, precedents are retrieved leave-one-out from the corresponding cleaned evaluation population. A precedent contains the previous content and its observed decision. Outcome labels are usually semantic, but HateModerate also includes an opaque RAD control in which policy labels are replaced with arbitrary consistent identifiers.

The rule-inventory variants separate information effects from answer-space effects. \emph{All} exposes the complete applicable ruleset. \emph{Represented} exposes rules supported by the designated precedent corpus; in the tables this appears as corpus-represented or train-represented depending on the source corpus. \emph{Retrieved} exposes only rules appearing among the retrieved violating precedents. Thus, \emph{General-all} versus \emph{NAD-all} isolates the effect of adding precedents while holding the complete answer space fixed, whereas \emph{General-retrieved} versus semantic \emph{RAD} isolates the effect of adding written rule descriptions under the retrieved candidate set. For harmful-content detection there is no independent written ruleset, so the matched augmentation contrast is \emph{Direct} versus semantic \emph{RAD}.

\subsection{SOM Decision Interfaces}

We evaluate two SOM implementations, Jev and Laya. Information settings are kept conceptually distinct from the typed interface used to obtain a decision. Depending on the experiment, we employ three SOM interfaces.

\begin{itemize}
    \item \textbf{Choice} selects one outcome from a finite candidate set. It returns the selected option, a probability distribution over the available options, and a separate confidence value. Choice is required for exact policy selection and is also evaluated as a binary interface on OLID, HateXplain, and ModerationBench.

    \item \textbf{Noul} evaluates a binary proposition and returns a probability in $[0,1]$, without a separate confidence field. We use it for binary harmful-content detection and for category-wise toxicity questions.

    \item \textbf{Score} positions an input on an ordinal scale specified by two to ten textual anchors and can return values between adjacent anchors. We use this interface in the exploratory comparison with Detoxify category-score vectors.
\end{itemize}

Thus, the information condition determines \emph{what evidence} is supplied to the model, whereas the SOM interface determines \emph{how the decision is elicited}. Comparisons intended to measure augmentation effects always retain the same interface.

\subsection{Cross-System Controls}

Jev and Laya differ in their available context budgets and provider-specific input serialization. We therefore construct shared request content before either system is invoked whenever the two implementations are compared. The target state, instructions, candidate answers, retrieved precedents, and deterministic text shortening are fixed independently of the evaluated SOM.

For PluRule and HateModerate, we retain the original complete Jev representation as \emph{Jev-full} and additionally evaluate \emph{Jev-compact}, which receives the same Laya-compatible content as Laya. All direct Jev--Laya comparisons in these experiments use Jev-compact. The OLID, HateXplain, and ModerationBench experiments use shared representations directly and therefore do not require separate full and compact Jev conditions. 

\subsection{Evaluation Strategy}
Our evaluation separates four aspects of moderation performance: correctness of the final moderation outcome, binary detection quality, discrimination independently of the decision rule, and the operational usefulness of model uncertainty.

For \emph{policy-violation detection}, the primary metric is \emph{Exact-Match Accuracy} (EMA), which requires the model either to identify the annotated violated rule or to correctly select \emph{No rules broken}. Because the same EMA can arise from substantially different moderation behaviors, we additionally examine class-conditional diagnostics. On PluRule, where each violating target is paired with a compliant target, we report \emph{Violating-Rule Recall} (VRR), measuring the fraction of violating targets for which the exact annotated rule is recovered, and \emph{No-Violation Recall} (NVR), measuring correct compliant decisions. On HateModerate, whose cleaned evaluation population is substantially skewed toward violating examples, we instead supplement EMA and VRR with binary violation accuracy and violation recall. These measures allow us to distinguish successful policy selection from operating points dominated by either compliance predictions or indiscriminate flagging. In all policy-violation experiments, the primary retrieval contrast is \emph{General-all versus NAD-all}, which adds retrieved precedents while holding the complete policy inventory and answer space fixed.
For \emph{harmful-content detection}, we distinguish operating-point classification performance from the quality of the underlying decision signal. Accuracy and F1 summarize operating-point performance, while AUROC measures the ability of the returned score to discriminate harmful from non-harmful content independently of the decision rule. For OLID and HateXplain, the primary matched augmentation analysis uses Noul, with AUROC as the primary endpoint; Choice provides a complementary fixed-decision comparison with Detoxify. On ModerationBench, AUROC is evaluated under both Choice and Noul. Accuracy and F1 characterize the resulting operating point.

We evaluate uncertainty separately from predictive performance. For binary decision probabilities, probability quality is summarized using \emph{predictive Brier score}, computed against the class label. For explicit Choice confidence, \emph{confidence Brier score} is computed against exact choice correctness, and correctness AUC measures whether confidence ranks correct decisions above errors. To assess whether confidence can support selective moderation, we additionally report the coverage and accuracy of predictions whose reported confidence satisfies $C\geq0.8$. These quantities respectively measure how much content could be handled automatically at that confidence threshold and how reliable those retained decisions would be.

For the exploratory comparison between SOM category scores and Detoxify outputs, we retain two complementary similarity measures: cosine similarity captures agreement in the relative profile across toxicity categories, whereas mean absolute error (MAE) captures disagreement in score magnitude.

Additional measures used for diagnostic or sensitivity analyses are reported in the Appendix. This includes alternative threshold metrics, calibration measures based on binning, held-out and representation-sensitivity checks, rule-cluster breakdowns, and results under alternative caption representations.
For the primary matched performance contrasts, and for selected secondary contrasts discussed explicitly below, we report 95\% bootstrap confidence intervals computed on the natural sampling unit of each benchmark. In PluRule, we resample complete discussion pairs so that the violating and compliant comments belonging to the same pair remain together. In HateModerate, OLID, HateXplain, and ModerationBench, we resample target items and preserve the pairing between the two compared conditions. For AUROC contrasts, each bootstrap replicate recomputes AUROC separately for the matched Direct and RAD predictions before taking their difference. Differences and confidence intervals are computed from unrounded estimates, while tables display rounded values. The reported intervals quantify individual contrasts and are not adjusted for multiple comparisons; they should therefore not be interpreted as simultaneous confidence statements.

\section{Experiments}
\label{sec:eval}
    
\subsection{Detection of Rule Violation}

We first evaluate policy-grounded moderation on PluRule and HateModerate. In both benchmarks, the model must distinguish violating from compliant content and, for violating targets, identify the applicable rule.

\begin{tcolorbox}[
    colback=gray!5,
    colframe=black!60,
    boxrule=0.6pt,
    arc=2pt,
    left=6pt,
    right=6pt,
    top=6pt,
    bottom=6pt,
    title=\textbf{Main Result},
    fonttitle=\normalsize
]
\begin{itemize}
    \item Performance is moderate overall: Jev consistently outperforms Laya in exact policy selection, but neither system is close to perfect.

    \item Retrieved precedents reliably improve Jev when the full rule inventory is fixed, with gains of +12.35, +5.20, and +22.20 EMA points across the three evaluations.

    \item Laya benefits much less from precedents and often behaves more like a broad violation detector than a precise policy selector.

    \item The precedent effect is strongest on HateModerate, substantial in high-support PluRule communities, and smaller on the PluRule test intersection.
\end{itemize}
\end{tcolorbox}

\noindent\textbf{PluRule moderation in large communities.}
We first consider the six PluRule communities with the largest numbers of complete discussion pairs: \texttt{changemyview} (1,583), \texttt{israelpalestine} (1,334), \texttt{excel} (1,182), \texttt{pokemontrades} (846), \texttt{ttc30} (739), and \texttt{tipofmytongue} (687). For each target, its complete discussion pair is excluded before retrieving the five nearest same-subreddit precedents or deriving precedent-supported rules. Appendix~\ref{app:six-community-heldout} repeats the analysis on official validation and test targets using training decisions only.

\begin{table}[t]
\centering
\scriptsize
\setlength{\tabcolsep}{1.8mm}
\caption{PluRule results in the six largest communities. Jev-compact and Laya receive the same compact request content. EMA is exact-match accuracy, VRR is violating-rule recall, and NVR is no-violation recall. Best and second-best values are bold and italicized, respectively.}
\label{tab:six-subreddit-transductive}
\begin{tabular}{llrrr}
\toprule
Model & Decision arm & EMA & VRR & NVR \\
\midrule
Jev-compact & NAD---all rules & 52.7 & 37.1 & 68.3 \\
Jev-compact & RAD---precedents only & \textbf{72.5} & \textit{63.9} & \textit{81.1} \\
Jev-compact & General---all rules & 65.1 & 62.3 & 67.8 \\
Jev-compact & General---retrieved rules & \textit{71.1} & \textbf{66.0} & 76.1 \\
\midrule
Laya & NAD---all rules & 38.1 & 1.6 & 74.5 \\
Laya & RAD---precedents only & 46.9 & 28.6 & 65.2 \\
Laya & General---all rules & 36.7 & 5.4 & 67.9 \\
Laya & General---retrieved rules & 48.9 & 10.2 & \textbf{87.6} \\
\bottomrule
\end{tabular}
\end{table}
Table~\ref{tab:six-subreddit-transductive} reports the controlled Jev-compact and Laya results.
The matched all-rule contrast shows a substantial precedent effect for Jev: General-all raises Jev-compact EMA by 12.35 points over NAD-all (95\% pair-bootstrap CI 11.62--13.09) and raises VRR from 37.1\% to 62.3\%. The Jev-full rendering gives the same qualitative result, with an 11.7-point EMA gain (95\% CI 11.0--12.4). Laya does not show the same response: General-all reduces EMA by 1.4 points (95\% CI $-2.1$ to $-0.7$). Restricted RAD-style conditions operate over a smaller answer space and exhibit strong VRR--NVR trade-offs. Even with this restricted answer space held fixed, written rule information can substantially shift the operating point: for Laya, moving from RAD to General-retrieved raises NVR from 65.2\% to 87.6\% while lowering VRR from 28.6\% to 10.2\%.

\noindent\textbf{PluRule test-intersection comparison.}
We next evaluate the 2,829-pair PluRule test intersection. Five discussion pairs could not be hydrated, leaving 5,658 target decisions, corresponding to 99.82\% of the released test set. Unlike the preceding high-support setting, retrieval is restricted to the released training split: for each test target, RAD and General retrieve the five most similar training comments from the same subreddit, with at most one precedent per training pair. Validation and test labels are never used as precedent sources.
Table~\ref{tab:full-results} reports the principal SOM conditions together with the published PluRule reference results. The published systems are included as external reference points (see the original paper~\cite{plurule2026}).

\begin{table}[t]
\centering
\scriptsize
\setlength{\tabcolsep}{1.4mm}
\caption{Results on the 2,829-pair rehydrated PluRule test intersection. Jev-compact and Laya receive the same compact request content. Published PluRule systems are shown as external reference points. Bold and italicized values identify the best and second-best SOM results, respectively.}
\label{tab:full-results}
\begin{tabular}{llrrr}
\toprule
Model & Decision arm & EMA & VRR & NVR \\
\midrule
No-moderation baseline & Always no rules broken & 50.0 & 0.0 & 100.0 \\
Qwen3-VL 8B Instruct & Comment Only & 51.0 & -- & -- \\
GPT-5.2 Low & Comment Only & 54.1 & 40.1 & 68.1 \\
GPT-5.2 High & Comment Only & 55.0 & 40.9 & 69.2 \\
GPT-5.2 High & Best full context & 57.7 & -- & -- \\
\midrule
Jev-compact & NAD---all rules & 48.4 & \textit{35.3} & 61.4 \\
Jev-compact & RAD---precedents only & \textbf{58.3} & 26.5 & \textit{90.0} \\
Jev-compact & General---all rules & 53.6 & \textbf{43.9} & 63.2 \\
Jev-compact & General---retrieved rules & \textit{58.2} & 28.7 & 87.6 \\
\midrule
Laya & NAD---all rules & 35.8 & 3.9 & 67.7 \\
Laya & RAD---precedents only & 47.3 & 15.9 & 78.6 \\
Laya & General---all rules & 37.1 & 6.3 & 67.9 \\
Laya & General---retrieved rules & 49.3 & 6.9 & \textbf{91.7} \\
\bottomrule
\end{tabular}
\end{table}

The test intersection reproduces the same pattern at smaller magnitude. Jev-compact gains 5.20 EMA points under General-all versus NAD-all (95\% CI 4.42--5.99), with VRR rising from 35.3\% to 43.9\%; Jev-full gains 4.4 points (95\% CI 3.7--5.2). Laya gains only 1.4 points (95\% CI 0.5--2.2). Restricted conditions again inflate raw EMA by favoring compliance: Jev-compact RAD reaches 58.3\% EMA but only 26.5\% VRR, and Laya's best EMA, 49.3\%, coincides with 91.7\% NVR and 6.9\% VRR. 

\noindent\textbf{Facebook policy moderation.}
HateModerate provides a complementary setting in which every target is evaluated against a fixed platform-wide taxonomy of 41 hate-speech policies. After grouping exact duplicates within each released split, removing two contradictorily annotated sentences, and excluding 175 multi-policy positives that cannot be represented by the single-choice interface, the corpus contains 3,309 training and 3,302 test sentences. The 114 sentences appearing in both released splits are retained as test targets but excluded from retrieval.

All 41 policies occur in the training set, so restricting the complete inventory to train-represented policies does not materially change the candidate space; those redundant conditions are therefore omitted here. We additionally compare two RAD representations using the same five retrieved training examples: \emph{RAD-opaque} replaces policy names with consistent arbitrary identifiers, whereas \emph{RAD-semantic} exposes the human-readable policy labels. Neither condition receives written policy descriptions. General-retrieved retains the same precedent-derived candidate set as semantic RAD but adds the corresponding written policy information.

Because HateModerate is strongly skewed toward violating targets, we supplement EMA and VRR with binary accuracy and violation recall. Recall is particularly informative here for distinguishing genuine policy selection from indiscriminate flagging.

\begin{table}[t]
\centering
\scriptsize
\setlength{\tabcolsep}{1.5mm}
\caption{HateModerate results on 3,302 cleaned test sentences. Jev-compact and Laya receive the same compact request content. Acc. and Rec. refer to the binary violation decision; EMA and VRR evaluate the exact moderation outcome. Best and second-best values are bold and italicized, respectively.}
\label{tab:hatemoderate}
\begin{tabular}{llrrrr}
\toprule
Model & Decision arm & Acc. & Rec. & EMA & VRR \\
\midrule
Jev-compact & NAD---all labels & 71.2 & 98.2 & 29.4 & 27.6 \\
Jev-compact & RAD---opaque & \textbf{83.3} & 92.6 & \textbf{52.5} & 40.7 \\
Jev-compact & RAD---semantic & \textit{78.2} & 90.3 & 50.9 & \textit{44.1} \\
Jev-compact & General---all labels & 76.6 & 96.5 & \textit{51.6} & \textbf{54.3} \\
Jev-compact & General---retrieved labels & 78.1 & 90.3 & 50.8 & \textit{44.1} \\
\midrule
Laya & NAD---all labels & 59.3 & \textbf{100.0} & 6.2 & 10.4 \\
Laya & RAD---opaque & 60.4 & 87.1 & 24.6 & 26.8 \\
Laya & RAD---semantic & 62.9 & 97.4 & 22.6 & 29.4 \\
Laya & General---all labels & 59.6 & \textit{99.7} & 15.5 & 25.4 \\
Laya & General---retrieved labels & 62.9 & 97.4 & 22.6 & 29.4 \\
\bottomrule
\end{tabular}
\end{table}

HateModerate gives the strongest matched precedent effect. General-all raises Jev-compact EMA by 22.20 points over NAD-all (95\% target-bootstrap CI 20.56--23.83) and nearly doubles VRR from 27.6\% to 54.3\%; Jev-full again preserves the direction of the result, with a 16.9-point EMA gain (95\% CI 15.4--18.4). Laya also improves under General-all, but from a much lower exact-selection baseline: its 100\% binary recall under NAD-all coexists with only 6.2\% EMA and 10.4\% VRR, indicating broad flagging rather than reliable policy identification.

The semantic--opaque RAD comparison shows that human-readable policy labels change the operating point without consistently improving exact policy application. For Jev-compact, semantic labels reduce binary accuracy by 5.06 points (95\% CI $-6.21$ to $-3.91$) and EMA by 1.64 points (95\% CI $-2.82$ to $-0.45$). For Laya, they raise binary accuracy by 2.5 points (95\% CI 1.4--3.7) but reduce EMA by 2.0 points (95\% CI $-3.3$ to $-0.7$).

\subsection{Detection of Harmful Content}
We next evaluate whether the behavior observed in policy-grounded moderation extends to harmful-content detection, where no explicit written policy is supplied. We consider OLID and HateXplain as text-only offensive-content benchmarks and ModerationBench as a broader \emph{Safe}/\emph{Unsafe} moderation task over Bluesky content. As described in Section~\ref{sec:model}, all cross-system comparisons use a shared input representation constructed before either SOM is invoked. In this setting, \emph{Direct} provides only the target and decision question, whereas \emph{RAD-semantic} additionally provides five semantically similar labeled precedents.

\begin{tcolorbox}[
    colback=gray!5,
    colframe=black!60,
    boxrule=0.6pt,
    arc=2pt,
    left=6pt,
    right=6pt,
    top=6pt,
    bottom=6pt,
    title=\textbf{Main Results},
    fonttitle=\normalsize
]
\begin{itemize}
    \item Jev shows competitive harmful-content detection, matching or exceeding Detoxify on F1 and clearly outperforming Laya across most settings.

    \item Retrieved precedents generally improve Jev's decision signal, increasing AUROC on OLID, HateXplain, and ModerationBench, although gains in threshold-dependent metrics are not universal.

    \item Laya is substantially less consistent: retrieval degrades discrimination on OLID and HateXplain but improves it on ModerationBench.

    \item ModerationBench shows the clearest cross-system retrieval benefit: precedents improve discrimination for both Jev and Laya, unlike OLID and HateXplain.
\end{itemize}
\end{tcolorbox}

\noindent\textbf{Text-only offensive-content detection.}
After removing exact duplicates and contradictory annotations, the evaluation populations contain 14,046 OLID posts and 14,123 HateXplain posts. We first compare the binary Choice formulation of Jev and Laya with the five Detoxify variants~\cite{hanu2020detoxify}. 

Table~\ref{tab:binary-detoxify} reports Accuracy and F1. On OLID, Detoxify obtains the highest accuracy, with the Multilingual variant reaching 81.0\%. Jev Direct instead attains the highest F1, 66.1\%, closely followed by Detoxify-Multilingual at 65.7\%. Adding precedents does not improve Jev's Choice result on this dataset: RAD-semantic lowers accuracy from 74.4\% to 73.5\% and F1 from 66.1\% to 63.3\%. Laya shows a larger deterioration in accuracy, from 72.6\% to 59.6\%, despite a modest increase in F1.
The pattern differs on HateXplain. Jev RAD-semantic reaches 66.1\% accuracy and 65.9\% F1, above its Direct condition and the external Detoxify reference values shown in the table. Jev Direct already obtains an F1 of 65.5\%, compared with at most 57.0\% for Detoxify. Laya remains less consistent: retrieval reduces accuracy from 59.6\% to 56.0\%, although F1 rises substantially. Thus, the operating-point results already indicate that the effect of precedents is model- and dataset-dependent.

\begin{table}[t]
\centering
\scriptsize
\setlength{\tabcolsep}{1.8mm}
\caption{Binary Choice offensiveness detection on the cleaned OLID and HateXplain populations. Accuracy and F1 are percentages. Detoxify predicts offensive when at least one category score is strictly greater than 0.5. Detoxify variants are external task-specific references rather than input-matched competitors.}
\label{tab:binary-detoxify}
\resizebox{0.72\textwidth}{!}{%
\begin{tabular}{llrr}
\toprule
System & Information / variant & Acc. & F1 \\
\midrule
\multicolumn{4}{l}{\textbf{Panel A: OLID (14,046 posts)}} \\
\midrule
Jev & Direct & 74.4 & 66.1 \\
Jev & RAD-semantic & 73.5 & 63.3 \\
Laya & Direct & 72.6 & 50.8 \\
Laya & RAD-semantic & 59.6 & 56.6 \\
\midrule
Detoxify & Original & 79.7 & 64.3 \\
Detoxify & Original-small & 79.5 & 62.7 \\
Detoxify & Unbiased & 80.6 & 64.4 \\
Detoxify & Unbiased-small & 80.0 & 61.6 \\
Detoxify & Multilingual & 81.0 & 65.7 \\
\midrule
\multicolumn{4}{l}{\textbf{Panel B: HateXplain offensive/normal subset (14,123 posts)}} \\
\midrule
Jev & Direct & 59.8 & 65.5 \\
Jev & RAD-semantic & 66.1 & 65.9 \\
Laya & Direct & 59.6 & 11.3 \\
Laya & RAD-semantic & 56.0 & 57.7 \\
\midrule
Detoxify & Original & 58.2 & 57.0 \\
Detoxify & Original-small & 59.5 & 56.6 \\
Detoxify & Unbiased & 57.9 & 56.3 \\
Detoxify & Unbiased-small & 58.4 & 54.3 \\
Detoxify & Multilingual & 57.6 & 56.7 \\
\bottomrule
\end{tabular}%
}
\end{table}

\noindent\textbf{Effect of precedent augmentation.}
We isolate the contribution of retrieved precedents more directly through the Noul interface. For each target, Direct and RAD-semantic ask exactly the same binary offensiveness question and differ only in whether the five retrieved labeled examples are supplied.

Table~\ref{tab:binary-single-noul} shows a consistent positive retrieval effect for Jev. On OLID, AUROC increases from .789 to .816 ($\Delta=.027$, 95\% paired-bootstrap CI .023--.032), while accuracy rises from 69.7\% to 74.0\% and F1 from 62.3\% to 64.6\%. HateXplain shows a smaller but still positive change in discrimination: AUROC rises from .749 to .755 ($\Delta=.005$, 95\% CI .001--.009), accompanied by a 7.0-point accuracy increase and an F1 increase from 64.5\% to 66.3\%.
Laya responds in the opposite direction on both datasets. On OLID, retrieval reduces AUROC from .800 to .702 ($\Delta=-.097$, 95\% CI $-.106$ to $-.089$) and accuracy from 74.5\% to 52.9\%. On HateXplain, AUROC falls from .604 to .515 ($\Delta=-.089$, 95\% CI $-.099$ to $-.078$) and accuracy from 56.1\% to 43.3\%, even though F1 increases from 57.3\% to 59.3\%. The latter combination is important: improved F1 at a particular threshold coexists with markedly worse ranking and overall accuracy. Diagnostic class-propensity results reported in the Appendix confirm that this behavior is associated with substantially more positive predictions under RAD rather than with improved discrimination.

\begin{table}[t]
\centering
\scriptsize
\setlength{\tabcolsep}{1.8mm}
\caption{Direct and semantic-RAD Noul offensiveness detection. Binary predictions use the fixed threshold $p(\mathrm{Offensive})>0.5$. AUROC is the primary retrieval endpoint; Accuracy and F1 characterize the resulting operating point. Accuracy and F1 are percentages. Best and second-best values within each dataset are bold and italicized, respectively.}
\label{tab:binary-single-noul}
\resizebox{0.72\textwidth}{!}{%
\begin{tabular}{llrrr}
\toprule
Model & Information & Acc. & F1 & AUROC \\
\midrule
\multicolumn{5}{l}{\textbf{Panel A: OLID (14,046 posts)}} \\
\midrule
Jev & Direct & 69.7 & \textit{62.3} & .789 \\
Jev & RAD-semantic & \textit{74.0} & \textbf{64.6} & \textbf{.816} \\
Laya & Direct & \textbf{74.5} & 62.0 & \textit{.800} \\
Laya & RAD-semantic & 52.9 & 56.1 & .702 \\
\midrule
\multicolumn{5}{l}{\textbf{Panel B: HateXplain offensive/normal subset (14,123 posts)}} \\
\midrule
Jev & Direct & \textit{56.3} & \textit{64.5} & \textit{.749} \\
Jev & RAD-semantic & \textbf{63.3} & \textbf{66.3} & \textbf{.755} \\
Laya & Direct & 56.1 & 57.3 & .604 \\
Laya & RAD-semantic & 43.3 & 59.3 & .515 \\
\bottomrule
\end{tabular}%
}
\end{table}

\noindent\textbf{Moderation of Bluesky content.}
We finally evaluate harmful-content detection on the text-based ModerationBench adaptation. The cleaned population contains 3,689 posts, comprising 2,584 \emph{Safe} and 1,105 \emph{Unsafe} examples. Each target is evaluated through both Choice and Noul. Direct uses the target alone, while RAD-semantic adds the five closest leave-one-out precedents with their human-readable \emph{Safe}/\emph{Unsafe} decisions.

Table~\ref{tab:moderationbench-binary} reports the pooled metrics retained by our evaluation strategy. Unlike OLID and HateXplain, retrieval improves Jev consistently under both interfaces. With Choice, AUROC increases from .893 to .936 ($\Delta=.043$, 95\% paired-bootstrap CI .034--.053), accuracy from 84.0\% to 86.7\%, and F1 from 73.2\% to 79.1\%, while predictive Brier improves from .121 to .097. The Noul results are similar: AUROC increases from .902 to .940 ($\Delta=.038$, 95\% CI .031--.046), accuracy reaches 87.4\%, F1 reaches 79.6\%, and predictive Brier decreases from .112 to .094. Thus, for Jev, the retrieved examples improve both ranking and operating-point moderation while also improving the quality of the returned probabilities.

Retrieval also improves Laya's discrimination on ModerationBench, but the operating-point consequences depend strongly on the interface. Choice AUROC rises from .521 to .661 ($\Delta=.141$, 95\% CI .116--.165), accompanied by a large increase in accuracy from 46.5\% to 66.0\% and an improvement in predictive Brier from .344 to .255. Noul likewise improves in AUROC, from .658 to .723 ($\Delta=.065$, 95\% CI .046--.084), but accuracy falls from 62.2\% to 50.5\% and predictive Brier worsens from .313 to .447. The same retrieval signal can therefore improve ranking while producing a poor operating point, reinforcing the need to distinguish AUROC from threshold-dependent metrics.

\begin{table}[t]
\centering
\scriptsize
\setlength{\tabcolsep}{1.6mm}
\caption{Pooled results on the cleaned ModerationBench adaptation ($n=3{,}689$). C and N denote Choice and Noul. Accuracy and F1 are percentages; AUROC and predictive Brier are reported on $[0,1]$. Choice predictive Brier uses the probability assigned to the \emph{Unsafe} class rather than the separate Choice confidence field. Higher is better except for Brier. Best and second-best values are bold and italicized, respectively.}
\label{tab:moderationbench-binary}
\resizebox{0.76\textwidth}{!}{%
\begin{tabular}{lllrrrr}
\toprule
Model & Information & Output & Acc. & F1 & AUROC & Brier \\
\midrule
Jev & Direct & C & 84.0 & 73.2 & .893 & .121 \\
Jev & RAD-semantic & C & \textit{86.7} & \textit{79.1} & \textit{.936} & \textit{.097} \\
Jev & Direct & N & 84.4 & 71.8 & .902 & .112 \\
Jev & RAD-semantic & N & \textbf{87.4} & \textbf{79.6} & \textbf{.940} & \textbf{.094} \\
\midrule
Laya & Direct & C & 46.5 & 42.4 & .521 & .344 \\
Laya & RAD-semantic & C & 66.0 & 46.3 & .661 & .255 \\
Laya & Direct & N & 62.2 & 46.7 & .658 & .313 \\
Laya & RAD-semantic & N & 50.5 & 53.8 & .723 & .447 \\
\bottomrule
\end{tabular}%
}
\end{table}

\subsection{Confidence and Selective Prediction}
Having evaluated predictive performance and the effect of precedents, we finally examine whether the confidence returned by the Choice interface can identify reliable decisions and support selective moderation.
The Choice interface returns a confidence value separately from its option probabilities. We evaluate this quantity against exact choice correctness: exact rule selection or a correct \emph{No rules broken} decision in policy-selection tasks, and the annotated binary outcome in harmful-content tasks. Deterministic single-option cases that do not invoke the model are excluded from confidence analysis; consequently, Exact Acc. may differ from the population-level EMA reported above, and coverage is computed over model-inferred decisions only. Table~\ref{tab:confidence-summary} mixes primary policy conditions and selected diagnostic conditions: the matched General-all policy condition for PluRule, the augmented condition with the highest population-level EMA for HateModerate, and semantic RAD for harmful-content detection. Additional held-out Choice-confidence diagnostics and predictive-probability sensitivity analyses are reported in the Appendix.

\begin{table}[t]
\centering
\scriptsize
\setlength{\tabcolsep}{1.2mm}
\caption{Summary of Choice-confidence behavior in representative conditions. Exact Acc., coverage, and accepted accuracy are percentages. Confidence Brier and correctness AUC are on $[0,1]$.}
\label{tab:confidence-summary}
\resizebox{0.88\textwidth}{!}{%
\begin{tabular}{lllrrrrr}
\toprule
Benchmark & Model & Condition & Exact Acc. & Brier & AUC & Cov$_{.8}$ & Acc$_{.8}$ \\
\midrule
Six-community PluRule & Jev-compact & General-all & 65.1 & .211 & .700 & 45.0 & 79.5 \\
Six-community PluRule & Laya & General-all & 36.7 & .242 & .593 & 2.0 & 49.6 \\
PluRule test intersection & Jev-compact & General-all & 53.6 & .264 & .663 & 41.0 & 67.6 \\
PluRule test intersection & Laya & General-all & 37.1 & .248 & .589 & 4.4 & 56.0 \\
HateModerate & Jev-compact & RAD-opaque & 51.0 & .223 & .749 & 35.5 & 75.8 \\
HateModerate & Laya & RAD-opaque & 22.0 & .214 & .570 & 5.2 & 27.9 \\
OLID & Jev & RAD-semantic & 73.5 & .203 & .718 & 54.6 & 85.5 \\
HateXplain & Jev & RAD-semantic & 66.1 & .255 & .628 & 48.3 & 74.9 \\
ModerationBench & Jev & RAD-semantic & 86.7 & .111 & .823 & 71.9 & 94.3 \\
ModerationBench & Laya & RAD-semantic & 66.0 & .346 & .560 & 19.2 & 63.4 \\
\bottomrule
\end{tabular}%
}
\end{table}

Jev confidence is useful for selective review across both policy-selection and harmful-content settings. High-confidence subsets are consistently more accurate than the unfiltered predictions and often retain a substantial fraction of cases, especially on ModerationBench. Laya's confidence is less reliable for automation: even when confidence Brier is competitive, as in HateModerate RAD-opaque, correctness AUC and high-confidence coverage remain weak. This separation between probability quality, correctness ranking, and accepted-set accuracy reinforces the need to report calibration and selective prediction together rather than treating confidence as a single scalar guarantee.

\section{Conclusion}
\label{sec:conclusion}

This paper evaluated whether System One Models can serve as structured decision models for online moderation. Across five benchmarks, Jev is a strong moderation system, performing competitively with specialized or published reference systems in several policy-grounded and harmful-content settings. It also benefits from retrieved precedents in the main matched contrasts: precedents improve exact policy selection on PluRule and HateModerate and improve harmful-content discrimination in the Noul and ModerationBench experiments. Laya, the main open source alternative, behaves differently, with retrieval more often changing the operating point than improving the underlying decision signal. The broader lesson is methodological as much as model-specific: moderation benchmarks should separate raw performance, policy information, precedent information, answer-space restriction, discrimination, and confidence. Typed probabilities make these distinctions easier to measure, but they do not by themselves guarantee policy adaptation or reliable automation.

\noindent\textbf{Limitations and future work}
This work is not exempt from limitations. We evaluate only two SOM implementations, so differences between Jev and Laya should not be generalized to all structured-output models, nor attributed solely to architecture. Finally, retrieved precedents may reproduce historical inconsistency or bias; provenance makes that risk visible, but does not eliminate it.
Future evaluations should test additional SOMs and generative baselines under the same controlled information conditions, vary the number and diversity of retrieved precedents, and study \textit{when} a model should rely on retrieved cases versus the complete written policy. Moderation settings also require support for multiple simultaneous violations, multilingual and community-specific deployment studies, drift monitoring, group-conditioned error analysis, and calibration methods tied to human review capacity. In practice, confidence should become part of an audited review policy, evaluated jointly with coverage, appeal outcomes, and class-specific error costs.

\appendix

\section{Additional Rule-Violation Diagnostics}
\label{app:rule-violation-diagnostics}

The main text reports the metric subset defined in Section~\ref{sec:model} and emphasizes the matched information contrasts used to answer the research questions. This appendix provides additional aggregate diagnostics for the policy-violation experiments. In particular, it restores the binary violation metrics, the intermediate represented-rule conditions, and the Jev-full representation-sensitivity runs.

\subsection{Six-Community PluRule Diagnostics}
\label{app:six-community-diagnostics}

Table~\ref{tab:appendix-six-subreddit-transductive} expands the transductive six-community analysis reported in the main text. In addition to EMA, VRR, and NVR, it reports binary accuracy, precision, recall, and F1 for the decision of whether any rule was violated. It also includes the corpus-represented conditions and the original Jev-full rendering.

\begin{table}[p]
\centering
\scriptsize
\setlength{\tabcolsep}{0.5mm}
\caption{Complete PluRule diagnostics in the six largest communities. Jev-full uses the original complete rendering; Jev-compact and Laya receive the same compact request content. Every one of the 12,742 comments is classified once. Binary metrics treat any violation as positive. EMA requires either the exact annotated rule or the correct \emph{No rules broken} outcome.}
\label{tab:appendix-six-subreddit-transductive}
\resizebox{\textwidth}{!}{%
\begin{tabular}{llrrrrrrr}
\toprule
& & \multicolumn{4}{c}{Violation decision} & \multicolumn{3}{c}{Exact outcomes} \\
\cmidrule(lr){3-6}\cmidrule(l){7-9}
Model & Decision arm & Acc. & Prec. & Rec. & F1 & EMA & VRR & NVR \\
\midrule
Jev-full & NAD---all rules & 67.2 & 68.6 & 63.4 & 65.9 & 56.1 & 41.1 & 71.0 \\
Jev-full & NAD---corpus represented & 71.0 & 72.6 & 67.4 & 69.9 & 61.8 & 49.1 & 74.6 \\
Jev-full & RAD---precedents only & 74.9 & 78.7 & 68.5 & 73.2 & 72.8 & 64.2 & 81.4 \\
Jev-full & General---all rules & 73.6 & 73.3 & 74.3 & 73.8 & 67.8 & 62.6 & 73.0 \\
Jev-full & General---corpus represented & 75.1 & 74.6 & 76.2 & 75.4 & 69.9 & 65.7 & 74.1 \\
Jev-full & General---retrieved rules & 75.4 & 76.6 & 73.0 & 74.8 & 72.7 & 67.7 & 77.8 \\
\midrule
Jev-compact & NAD---all rules & 64.4 & 65.6 & 60.4 & 62.9 & 52.7 & 37.1 & 68.3 \\
Jev-compact & NAD---corpus represented & 69.4 & 70.5 & 66.6 & 68.5 & 59.3 & 46.5 & 72.2 \\
Jev-compact & RAD---precedents only & 74.7 & 78.3 & 68.3 & 73.0 & 72.5 & 63.9 & 81.1 \\
Jev-compact & General---all rules & 71.1 & 69.8 & 74.4 & 72.0 & 65.1 & 62.3 & 67.8 \\
Jev-compact & General---corpus represented & 72.6 & 71.3 & 75.7 & 73.4 & 67.5 & 65.5 & 69.5 \\
Jev-compact & General---retrieved rules & 73.9 & 75.0 & 71.7 & 73.3 & 71.1 & 66.0 & 76.1 \\
\midrule
Laya & NAD---all rules & 49.9 & 49.8 & 25.2 & 33.5 & 38.1 & 1.6 & 74.5 \\
Laya & NAD---corpus represented & 49.2 & 47.6 & 14.9 & 22.7 & 42.7 & 1.9 & 83.6 \\
Laya & RAD---precedents only & 50.6 & 50.9 & 36.1 & 42.2 & 46.9 & 28.6 & 65.2 \\
Laya & General---all rules & 48.6 & 47.8 & 29.4 & 36.4 & 36.7 & 5.4 & 67.9 \\
Laya & General---corpus represented & 50.0 & 49.9 & 15.8 & 24.0 & 45.7 & 7.3 & 84.2 \\
Laya & General---retrieved rules & 50.3 & 51.2 & 13.1 & 20.8 & 48.9 & 10.2 & 87.6 \\
\bottomrule
\end{tabular}%
}
\end{table}

The binary metrics provide additional context for the exact-outcome results. Jev's General conditions increase both violation recall and exact-rule recovery relative to the corresponding normative conditions. Laya instead shows a stronger dependence on the candidate space: RAD increases its binary violation recall and VRR, whereas the restricted General conditions increasingly favor \emph{No rules broken}. The corpus-represented conditions exhibit the same pattern and do not alter the conclusions drawn from the simpler main-text comparison.

The Jev-full rows also provide a representation-sensitivity check. The matched General-all versus NAD-all EMA gain is 11.7 percentage points under the full rendering, close to the 12.35-point gain obtained with Jev-compact. Thus, the main precedent effect does not depend on the deterministic compacting required for the controlled Jev--Laya comparison.

\subsection{PluRule Aggregate Diagnostics}
\label{app:full-plurule-diagnostics}

Table~\ref{tab:appendix-full-plurule} restores the intermediate train-represented conditions and the Jev-full runs for the 2,829-pair rehydrated PluRule test intersection. The published systems are retained as descriptive external reference points. Traditional binary accuracy, precision, recall, and F1 were not reported for the published PluRule systems, so this comparison remains focused on exact outcomes.

\begin{table}[p]
\centering
\scriptsize
\setlength{\tabcolsep}{1.0mm}
\caption{Complete aggregate results on the 2,829-pair rehydrated PluRule test intersection. Jev-full uses the original complete rendering; Jev-compact and Laya receive the same compact request content. Published PluRule systems are descriptive external references.}
\label{tab:appendix-full-plurule}
\begin{tabular}{llrrr}
\toprule
Model & Decision arm & EMA & VRR & NVR \\
\midrule
No-moderation baseline & Always no rules broken & 50.0 & 0.0 & 100.0 \\
Qwen3-VL 8B Instruct & Comment Only & 51.0 & -- & -- \\
GPT-5.2 Low & Comment Only & 54.1 & 40.1 & 68.1 \\
GPT-5.2 High & Comment Only & 55.0 & 40.9 & 69.2 \\
GPT-5.2 High & Best full context & 57.7 & -- & -- \\
\midrule
Jev-full & NAD---all rules & 52.0 & 36.9 & 67.0 \\
Jev-full & NAD---train represented & 56.0 & 24.7 & 87.2 \\
Jev-full & RAD---precedents only & 58.9 & 26.4 & 91.4 \\
Jev-full & General---all rules & 56.4 & 43.9 & 68.9 \\
Jev-full & General---train represented & 57.9 & 29.6 & 86.2 \\
Jev-full & General---retrieved rules & 58.7 & 28.6 & 88.8 \\
\midrule
Jev-compact & NAD---all rules & 48.4 & 35.3 & 61.4 \\
Jev-compact & NAD---train represented & 54.7 & 24.1 & 85.3 \\
Jev-compact & RAD---precedents only & 58.3 & 26.5 & 90.0 \\
Jev-compact & General---all rules & 53.6 & 43.9 & 63.2 \\
Jev-compact & General---train represented & 57.5 & 29.8 & 85.1 \\
Jev-compact & General---retrieved rules & 58.2 & 28.7 & 87.6 \\
\midrule
Laya & NAD---all rules & 35.8 & 3.9 & 67.7 \\
Laya & NAD---train represented & 46.4 & 5.8 & 87.1 \\
Laya & RAD---precedents only & 47.3 & 15.9 & 78.6 \\
Laya & General---all rules & 37.1 & 6.3 & 67.9 \\
Laya & General---train represented & 47.4 & 5.2 & 89.5 \\
Laya & General---retrieved rules & 49.3 & 6.9 & 91.7 \\
\bottomrule
\end{tabular}
\end{table}

The train-represented conditions reinforce the answer-space effect discussed in the main text. Restricting the rule inventory generally raises EMA and NVR while reducing VRR, especially for Laya. For example, Laya increases from 35.8\% EMA under NAD-all to 46.4\% under NAD-train, despite VRR remaining only 5.8\%. Jev exhibits the same trade-off more moderately. These rows therefore support treating the complete-rule NAD-all--General-all comparison as the cleaner estimate of the contribution of retrieved precedents.

The Jev-full sensitivity analysis again preserves the principal matched effect: General-all increases EMA from 52.0\% to 56.4\%, compared with 48.4\% to 53.6\% under Jev-compact.

\subsection{HateModerate Detailed Diagnostics}
\label{app:hatemoderate-diagnostics}

The main text reports binary accuracy and recall together with EMA and VRR. Table~\ref{tab:appendix-hatemoderate-multiclass} restores positive-class precision and F1, macro F1 over the 42 exact outcomes, the redundant train-represented conditions, and the Jev-full sensitivity runs.

\begin{table}[p]
\centering
\scriptsize
\setlength{\tabcolsep}{0.5mm}
\caption{Complete HateModerate results on 3,302 cleaned test sentences. Binary metrics treat policy violation as positive. EMA evaluates the complete 42-way outcome space including \emph{No rules broken}; MF1 is macro F1 over these exact outcomes. Jev-full uses the original complete rendering, while Jev-compact and Laya receive the same compact content.}
\label{tab:appendix-hatemoderate-multiclass}
\resizebox{\textwidth}{!}{%
\begin{tabular}{lrrrrrrr}
\toprule
& \multicolumn{4}{c}{Violation decision} & \multicolumn{3}{c}{Exact outcomes} \\
\cmidrule(lr){2-5}\cmidrule(lr){6-8}
Decision arm & Acc. & Prec. & Rec. & F1 & EMA & MF1 & VRR \\
\midrule
Jev-full NAD---all labels & 69.9 & 66.6 & 98.4 & 79.5 & 28.4 & 24.9 & 28.5 \\
Jev-full NAD---train represented & 70.1 & 66.8 & 98.6 & 79.6 & 28.5 & 25.2 & 28.4 \\
Jev-full RAD---opaque & 83.5 & 80.6 & 94.9 & 87.2 & 51.2 & 40.1 & 40.4 \\
Jev-full RAD---semantic & 78.0 & 77.4 & 88.8 & 82.7 & 51.6 & 43.8 & 44.4 \\
Jev-full General---all labels & 72.2 & 68.4 & 98.4 & 80.7 & 45.3 & 47.2 & 53.0 \\
Jev-full General---train represented & 72.2 & 68.4 & 98.5 & 80.8 & 45.5 & 47.4 & 53.4 \\
Jev-full General---retrieved labels & 75.6 & 72.5 & 94.8 & 82.2 & 45.9 & 41.3 & 44.8 \\
\midrule
Jev-compact NAD---all labels & 71.2 & 67.8 & 98.2 & 80.2 & 29.4 & 22.9 & 27.6 \\
Jev-compact NAD---train represented & 71.6 & 68.0 & 98.3 & 80.4 & 29.8 & 23.4 & 27.6 \\
Jev-compact RAD---opaque & 83.3 & 81.7 & 92.6 & 86.8 & 52.5 & 40.9 & 40.7 \\
Jev-compact RAD---semantic & 78.2 & 77.0 & 90.3 & 83.1 & 50.9 & 43.1 & 44.1 \\
Jev-compact General---all labels & 76.6 & 72.9 & 96.5 & 83.0 & 51.6 & 49.4 & 54.3 \\
Jev-compact General---train represented & 76.3 & 72.5 & 96.5 & 82.8 & 51.2 & 49.3 & 54.2 \\
Jev-compact General---retrieved labels & 78.1 & 76.8 & 90.3 & 83.0 & 50.8 & 43.0 & 44.1 \\
\midrule
Laya NAD---all labels & 59.3 & 59.3 & 100.0 & 74.4 & 6.2 & 6.7 & 10.4 \\
Laya NAD---train represented & 59.3 & 59.3 & 100.0 & 74.4 & 6.2 & 6.7 & 10.4 \\
Laya RAD---opaque & 60.4 & 61.7 & 87.1 & 72.3 & 24.6 & 24.9 & 26.8 \\
Laya RAD---semantic & 62.9 & 61.9 & 97.4 & 75.7 & 22.6 & 24.3 & 29.4 \\
Laya General---all labels & 59.6 & 59.5 & 99.7 & 74.5 & 15.5 & 20.5 & 25.4 \\
Laya General---train represented & 59.6 & 59.5 & 99.7 & 74.5 & 15.5 & 20.5 & 25.4 \\
Laya General---retrieved labels & 62.9 & 61.9 & 97.4 & 75.7 & 22.6 & 24.3 & 29.4 \\
\bottomrule
\end{tabular}%
}
\end{table}

The additional binary metrics make the operating-point differences explicit. In particular, Laya's 100\% violation recall under NAD-all coexists with only 6.2\% EMA and 10.4\% VRR. Conversely, Jev-compact RAD-opaque obtains the highest binary accuracy and F1, while General-all achieves the highest VRR and macro exact-outcome F1. The two criteria therefore capture different aspects of moderation quality.

Because every policy occurs in the HateModerate training set, the all-label and train-represented conditions have the same candidate coverage. They are nevertheless generated through separate request-construction paths, so the small differences in Table~\ref{tab:appendix-hatemoderate-multiclass} should be read as serialization and request-construction sensitivity rather than as an answer-coverage effect.

\section{Held-Out Six-Community Evaluation}
\label{app:six-community-heldout}

This appendix reports the train-sourced counterpart to the transductive six-community analysis in the main text. We run two separate held-out evaluations: one on 637 official test pairs (1,274 comments) and one on 638 official validation, or \emph{eval}, pairs (1,276 comments). In both evaluations, precedents and represented rules come only from the official training split and the same subreddit. Test and validation decisions never enter the information-source corpus. The six decision conditions and metrics follow the main-text analysis, but \emph{represented} here always means represented in training. These earlier Jev runs use the complete rendering and are labeled Jev-full. We did not repeat these ancillary splits with Jev-compact, so the Laya rows are retained as within-system references and are not used for cross-system comparisons.

\begin{table}[p]
\centering
\scriptsize
\setlength{\tabcolsep}{0.5mm}
\caption{Held-out PluRule results on the six largest communities. Jev-full uses the earlier complete rendering; no Jev-compact run was performed. Panel A classifies official test targets using only same-subreddit training examples as information sources; Panel B independently classifies official validation (eval) targets under the same train-sourced protocol. Binary metrics treat any violation as positive; EMA requires the correct rule or a correct no-violation decision.}
\label{tab:six-subreddit}
\resizebox{\textwidth}{!}{%
\begin{tabular}{llrrrrrrr}
\toprule
\multicolumn{9}{l}{\textbf{Panel A: test targets + training sources (637 pairs; 1,274 comments)}} \\
\midrule
& & \multicolumn{4}{c}{Violation decision} & \multicolumn{3}{c}{Exact outcomes} \\
\cmidrule(lr){3-6}\cmidrule(l){7-9}
Model & Decision arm & Acc. & Prec. & Rec. & F1 & EMA & VRR & NVR \\
\midrule
Jev-full & NAD---all rules & 66.6 & 67.5 & 63.9 & 65.6 & 56.1 & 43.0 & 69.2 \\
Jev-full & NAD---train represented & 71.4 & 72.9 & 68.0 & 70.3 & 63.1 & 51.5 & 74.7 \\
Jev-full & RAD---precedents only & 73.9 & 78.4 & 66.1 & 71.7 & 72.4 & 63.1 & 81.8 \\
Jev-full & General---all rules & 73.1 & 72.9 & 73.5 & 73.2 & 67.5 & 62.3 & 72.7 \\
Jev-full & General---train represented & 73.6 & 73.4 & 74.1 & 73.8 & 69.2 & 65.1 & 73.2 \\
Jev-full & General---retrieved rules & 74.3 & 76.7 & 69.9 & 73.1 & 72.6 & 66.4 & 78.8 \\
\midrule
Laya & NAD---all rules & 49.5 & 48.8 & 22.6 & 30.9 & 38.9 & 1.4 & 76.3 \\
Laya & NAD---train represented & 49.8 & 49.3 & 17.7 & 26.1 & 42.8 & 3.8 & 81.8 \\
Laya & RAD---precedents only & 50.5 & 50.7 & 35.6 & 41.8 & 46.5 & 27.6 & 65.3 \\
Laya & General---all rules & 47.8 & 46.2 & 26.7 & 33.8 & 36.9 & 4.9 & 68.9 \\
Laya & General---train represented & 48.6 & 44.2 & 10.7 & 17.2 & 45.8 & 5.2 & 86.5 \\
Laya & General---retrieved rules & 51.1 & 54.9 & 12.2 & 20.0 & 50.2 & 10.5 & 90.0 \\
\midrule
\multicolumn{9}{l}{\textbf{Panel B: validation/eval targets + training sources (638 pairs; 1,276 comments)}} \\
\midrule
Jev-full & NAD---all rules & 66.5 & 68.5 & 61.0 & 64.5 & 55.2 & 38.4 & 71.9 \\
Jev-full & NAD---train represented & 71.6 & 73.8 & 66.8 & 70.1 & 61.8 & 47.3 & 76.3 \\
Jev-full & RAD---precedents only & 74.3 & 78.5 & 66.9 & 72.3 & 72.3 & 62.9 & 81.7 \\
Jev-full & General---all rules & 74.4 & 75.0 & 73.0 & 74.0 & 68.1 & 60.5 & 75.7 \\
Jev-full & General---train represented & 75.0 & 75.1 & 74.8 & 74.9 & 69.7 & 64.3 & 75.2 \\
Jev-full & General---retrieved rules & 74.2 & 75.7 & 71.3 & 73.4 & 71.2 & 65.2 & 77.1 \\
\midrule
Laya & NAD---all rules & 48.7 & 47.5 & 24.9 & 32.7 & 37.3 & 2.2 & 72.4 \\
Laya & NAD---train represented & 47.7 & 43.8 & 16.0 & 23.4 & 41.1 & 2.7 & 79.5 \\
Laya & RAD---precedents only & 50.6 & 50.9 & 37.5 & 43.1 & 46.6 & 29.5 & 63.8 \\
Laya & General---all rules & 50.2 & 50.3 & 31.2 & 38.5 & 38.0 & 6.9 & 69.1 \\
Laya & General---train represented & 49.6 & 48.5 & 12.7 & 20.1 & 46.2 & 6.0 & 86.5 \\
Laya & General---retrieved rules & 51.0 & 53.7 & 14.9 & 23.3 & 49.5 & 11.8 & 87.1 \\
\bottomrule
\end{tabular}%
}
\end{table}

Table~\ref{tab:six-subreddit} reports the two held-out experiments separately. For Jev-full, the matched precedent effect recurs on both splits: General-all raises EMA over NAD-all from 56.1\% to 67.5\% on test and from 55.2\% to 68.1\% on validation. Laya's within-system ordering is also similar across splits, with RAD increasing VRR and General-retrieved favoring no-violation decisions. Because no Jev-compact counterpart was run, these rows support held-out robustness within each representation but do not enter the controlled Jev--Laya comparison.

\subsection{Held-Out Confidence and Calibration}

The Choice interface returns an explicit confidence value for each model inference. Table~\ref{tab:confidence-six} evaluates it as the probability that the selected multiclass outcome is exactly correct. Deterministic one-option cases are excluded because they invoke no model. Confidence Brier summarizes probability quality, ECE measures bin-based calibration, AUC measures whether confidence ranks correct decisions above errors, and Cov$_{.8}$ and Acc$_{.8}$ report selective coverage and accuracy at confidence $C\geq0.8$.

\begin{table}[p]
\centering
\scriptsize
\setlength{\tabcolsep}{0.5mm}
\caption{Confidence and exact-choice calibration on the combined validation and test predictions for the six largest subreddits (1,275 pairs before inference-only filtering). Jev denotes Jev-full; no Jev-compact run was performed. Det. is the percentage of all decisions excluded as deterministic; Cov$_{.8}$ and Acc$_{.8}$ are coverage and accuracy at $C\geq0.8$. Brier and AUC are on $[0,1]$; other performance values are percentages.}
\label{tab:confidence-six}
\resizebox{\textwidth}{!}{%
\begin{tabular}{llrrrrrrrrrrrr}
\toprule
Model & Arm & Inf. & Det. & Exact Acc. & $C$ & $C_+$ & $C_-$ & Gap & Brier & ECE & AUC & Cov$_{.8}$ & Acc$_{.8}$ \\
\midrule
Jev-full & NA & 2,550 & 0.0 & 55.6 & 67.9 & 73.0 & 61.5 & $+12.2$ & .246 & 12.8 & .664 & 34.4 & 70.8 \\
Jev-full & NT & 2,550 & 0.0 & 62.5 & 70.8 & 76.7 & 61.0 & $+8.3$ & .218 & 9.2 & .705 & 42.4 & 78.6 \\
Jev-full & R & 2,323 & 8.9 & 71.2 & 68.0 & 71.2 & 60.1 & $-3.2$ & .238 & 16.5 & .612 & 45.2 & 78.4 \\
Jev-full & GA & 2,550 & 0.0 & 67.8 & 73.2 & 77.7 & 63.8 & $+5.4$ & .204 & 8.4 & .692 & 46.6 & 81.2 \\
Jev-full & GT & 2,550 & 0.0 & 69.5 & 73.8 & 79.1 & 61.7 & $+4.4$ & .190 & 7.3 & .728 & 48.7 & 84.5 \\
Jev-full & GR & 2,323 & 8.9 & 70.7 & 68.4 & 74.1 & 54.7 & $-2.3$ & .211 & 12.6 & .694 & 47.4 & 82.2 \\
\midrule
Laya & NA & 2,550 & 0.0 & 38.1 & 29.8 & 30.2 & 29.5 & $-8.3$ & .272 & 16.7 & .516 & 2.4 & 45.2 \\
Laya & NT & 2,550 & 0.0 & 41.9 & 29.1 & 31.2 & 27.5 & $-12.8$ & .284 & 18.7 & .557 & 3.8 & 51.0 \\
Laya & R & 2,323 & 8.9 & 42.9 & 22.1 & 22.1 & 22.2 & $-20.7$ & .337 & 27.0 & .500 & 2.1 & 34.7 \\
Laya & GA & 2,550 & 0.0 & 37.5 & 30.2 & 34.1 & 27.8 & $-7.3$ & .241 & 10.3 & .598 & 1.9 & 59.2 \\
Laya & GT & 2,550 & 0.0 & 46.0 & 36.8 & 38.4 & 35.4 & $-9.2$ & .289 & 17.4 & .541 & 4.7 & 52.1 \\
Laya & GR & 2,323 & 8.9 & 46.5 & 33.7 & 33.5 & 34.0 & $-12.7$ & .340 & 26.5 & .489 & 7.3 & 50.0 \\
\bottomrule
\end{tabular}%
}
\end{table}

On the pooled six-community scope, Jev-full confidence contains meaningful information about correctness. All six AUCs exceed 0.61, with General-train reaching 0.728. Its accepted high-confidence decisions cover 48.7\% of inferred cases and are 84.5\% accurate, compared with 69.5\% before abstention. Precedents also improve calibration relative to their matched normative controls: General-all reduces the absolute confidence gap from 12.2 to 5.4 points versus NAD-all, while General-train reduces it from 8.3 to 4.4 points versus NAD-train.

Laya exhibits the opposite level bias: every arm is underconfident by 7.3 to 20.7 points. More importantly, confidence often fails to distinguish its correct from incorrect answers. RAD has virtually identical $C_+$ and $C_-$ and AUC 0.500; General-retrieved falls below chance at 0.489. General-all is the strongest Laya discriminator (AUC 0.598), but only 1.9\% of its decisions clear the 0.8 threshold. Laya's low confidence is therefore not, by itself, a reliable mechanism for deciding which cases to automate.

\clearpage
\section{PluRule Rule-Cluster Breakdowns}
\label{app:plurule-clusters}

The aggregate PluRule test result hides substantial differences across the benchmark's rule clusters. Table~\ref{tab:rule-clusters} reports exact-match accuracy by rule cluster. These earlier Jev results use Jev-full; no Jev-compact cluster run was performed, so the two systems are described separately. We abbreviate NAD-all, NAD-train, RAD, General-all, General-train, and General-retrieved as NA, NT, R, GA, GT, and GR, respectively; J denotes Jev-full and L denotes Laya. T-Cov is the exact annotated-rule coverage shared by NT and GT; R-Cov is the coverage shared by R and GR. NA and GA always have 100\% coverage.

The published reference column (GPT-F) is GPT-5.2 High with full context. It includes discussion, submission, user, and image context and is evaluated on all 2,834 test pairs. Our systems do not receive those four context levels and are evaluated on the 2,829-pair hydrated intersection. GPT-F is therefore a descriptive external reference rather than a controlled head-to-head condition. Its integer percentages are read from the paper's figures, whereas ours are computed from retained per-target predictions.

\begin{table}[p]
\centering
\scriptsize
\setlength{\tabcolsep}{0.5mm}
\caption{Test exact-match accuracy (\%) by rule cluster. GPT-F is the published GPT-5.2 High full-context result. T-Cov and R-Cov report train- and precedent-derived exact-rule coverage.}
\label{tab:rule-clusters}
\textit{(a) Published reference, coverage, and Jev-full}\par\smallskip
\resizebox{\textwidth}{!}{%
\begin{tabular}{lrrrrrrrrrr}
\toprule
Rule cluster & Pairs & GPT-F & T-Cov & R-Cov & J-NA & J-NT & J-R & J-GA & J-GT & J-GR \\
\midrule
Civility & 467 & 69 & 58.0 & 51.6 & 62.2 & 63.4 & 63.0 & 65.3 & 64.1 & 63.6 \\
Language & 25 & 66 & 60.0 & 24.0 & 58.0 & 62.0 & 56.0 & 60.0 & 64.0 & 56.0 \\
Self-promotion & 385 & 63 & 42.1 & 35.6 & 59.0 & 57.1 & 57.7 & 61.3 & 57.7 & 58.4 \\
Megathread & 24 & 62 & 20.8 & 12.5 & 58.3 & 52.1 & 50.0 & 60.4 & 54.2 & 52.1 \\
Doxxing & 24 & 60 & 25.0 & 16.7 & 43.8 & 52.1 & 54.2 & 45.8 & 56.2 & 56.2 \\
NSFW & 25 & 60 & 28.0 & 24.0 & 58.0 & 54.0 & 54.0 & 54.0 & 52.0 & 54.0 \\
Anti-piracy & 66 & 60 & 18.2 & 16.7 & 54.5 & 50.8 & 51.5 & 56.1 & 50.8 & 51.5 \\
Other & 713 & 59 & 69.6 & 62.7 & 51.4 & 56.2 & 64.5 & 60.7 & 61.9 & 65.1 \\
Spoilers & 105 & 56 & 35.2 & 31.4 & 48.6 & 52.9 & 54.3 & 50.5 & 53.8 & 54.8 \\
Attribution & 99 & 56 & 38.4 & 31.3 & 49.5 & 51.0 & 52.5 & 54.0 & 54.0 & 51.5 \\
No politics & 83 & 54 & 31.3 & 27.7 & 51.2 & 50.6 & 55.4 & 52.4 & 51.2 & 51.8 \\
Hate speech & 72 & 53 & 43.1 & 29.2 & 54.2 & 53.5 & 55.6 & 56.9 & 56.9 & 55.6 \\
Meta & 40 & 52 & 40.0 & 30.0 & 45.0 & 56.2 & 61.3 & 55.0 & 60.0 & 60.0 \\
Public discourse & 40 & 52 & 30.0 & 20.0 & 43.8 & 53.7 & 51.2 & 45.0 & 51.2 & 51.2 \\
On-topic & 34 & 51 & 55.9 & 44.1 & 38.2 & 48.5 & 52.9 & 47.1 & 50.0 & 51.5 \\
Context & 73 & 51 & 34.2 & 23.3 & 45.9 & 52.7 & 51.4 & 47.3 & 50.7 & 52.7 \\
Evidence-based & 46 & 47 & 71.7 & 65.2 & 54.3 & 63.0 & 62.0 & 55.4 & 60.9 & 59.8 \\
Flair & 140 & 46 & 70.0 & 66.4 & 47.1 & 56.8 & 63.6 & 52.9 & 57.9 & 58.9 \\
Anti-brigading & 38 & 46 & 23.7 & 18.4 & 36.8 & 46.1 & 48.7 & 39.5 & 46.1 & 46.1 \\
Meme ban & 31 & 45 & 16.1 & 16.1 & 40.3 & 51.6 & 53.2 & 45.2 & 53.2 & 51.6 \\
Titles & 88 & 45 & 18.2 & 14.8 & 35.8 & 49.4 & 51.7 & 38.1 & 50.0 & 49.4 \\
Reposts & 83 & 44 & 20.5 & 12.0 & 43.4 & 51.2 & 51.8 & 44.6 & 51.2 & 53.0 \\
Relevance & 33 & 44 & 21.2 & 9.1 & 43.9 & 54.5 & 50.0 & 45.5 & 53.0 & 50.0 \\
Karma & 31 & 44 & 45.2 & 41.9 & 32.3 & 53.2 & 53.2 & 37.1 & 51.6 & 53.2 \\
Low effort & 21 & 43 & 28.6 & 19.0 & 38.1 & 45.2 & 57.1 & 38.1 & 47.6 & 47.6 \\
Day-specific & 18 & 42 & 22.2 & 11.1 & 33.3 & 50.0 & 47.2 & 30.6 & 50.0 & 47.2 \\
Location-based & 25 & 30 & 20.0 & 8.0 & 38.0 & 48.0 & 52.0 & 40.0 & 50.0 & 50.0 \\
\bottomrule
\end{tabular}%
}
\medskip
\textit{(b) Laya}\par\smallskip
\begin{tabular}{lrrrrrr}
\toprule
Rule cluster & L-NA & L-NT & L-R & L-GA & L-GT & L-GR \\
\midrule
Civility & 34.8 & 47.1 & 46.5 & 36.7 & 48.6 & 51.2 \\
Language & 36.0 & 52.0 & 50.0 & 42.0 & 52.0 & 58.0 \\
Self-promotion & 37.5 & 45.6 & 46.8 & 39.2 & 47.8 & 49.2 \\
Megathread & 35.4 & 50.0 & 47.9 & 41.7 & 50.0 & 50.0 \\
Doxxing & 35.4 & 52.1 & 52.1 & 35.4 & 52.1 & 52.1 \\
NSFW & 34.0 & 52.0 & 50.0 & 36.0 & 48.0 & 44.0 \\
Anti-piracy & 33.3 & 47.7 & 47.0 & 35.6 & 50.0 & 50.0 \\
Other & 37.2 & 43.8 & 48.7 & 36.5 & 46.4 & 48.7 \\
Spoilers & 38.1 & 52.4 & 49.0 & 39.0 & 50.0 & 51.0 \\
Attribution & 32.8 & 47.5 & 47.5 & 32.8 & 49.0 & 47.5 \\
No politics & 39.2 & 50.6 & 47.6 & 36.7 & 48.2 & 51.8 \\
Hate speech & 31.9 & 47.2 & 45.8 & 35.4 & 45.1 & 46.5 \\
Meta & 31.2 & 42.5 & 47.5 & 36.2 & 45.0 & 47.5 \\
Public discourse & 33.7 & 46.2 & 47.5 & 37.5 & 48.8 & 48.8 \\
On-topic & 32.4 & 50.0 & 47.1 & 29.4 & 44.1 & 47.1 \\
Context & 33.6 & 43.2 & 41.8 & 36.3 & 43.8 & 45.2 \\
Evidence-based & 35.9 & 43.5 & 41.3 & 34.8 & 42.4 & 44.6 \\
Flair & 33.9 & 49.6 & 51.1 & 41.1 & 46.8 & 53.6 \\
Anti-brigading & 34.2 & 47.4 & 38.2 & 38.2 & 48.7 & 50.0 \\
Meme ban & 35.5 & 46.8 & 48.4 & 37.1 & 46.8 & 46.8 \\
Titles & 31.2 & 45.5 & 46.6 & 33.5 & 46.6 & 46.6 \\
Reposts & 36.7 & 48.2 & 44.6 & 36.7 & 47.6 & 47.6 \\
Relevance & 39.4 & 47.0 & 50.0 & 40.9 & 45.5 & 47.0 \\
Karma & 35.5 & 45.2 & 45.2 & 33.9 & 41.9 & 50.0 \\
Low effort & 38.1 & 45.2 & 40.5 & 42.9 & 47.6 & 47.6 \\
Day-specific & 47.2 & 52.8 & 50.0 & 50.0 & 52.8 & 52.8 \\
Location-based & 30.0 & 46.0 & 48.0 & 38.0 & 48.0 & 50.0 \\
\bottomrule
\end{tabular}
\end{table}

Across rule clusters, Jev-full NAD-all and General-all reproduce the published difficulty ordering with Spearman correlations of $\rho=0.845$ and $\rho=0.839$, respectively. General-all also reduces Jev-full's unweighted mean absolute difference from GPT-F, from 6.3 to 4.9 percentage points across rule clusters. This agreement is notable despite the additional context supplied to GPT-F, but it does not establish model equivalence. Laya's within-profile correlations range from $-0.103$ to $0.387$; because Jev-compact was not run for these slices, the correlation levels are not compared across systems.

\clearpage

\section{Additional Harmful-Content Diagnostics}
\label{app:harmful-content}

The main text reports the metric subset defined in Section~\ref{sec:model}: Accuracy and F1 for operating-point behavior, AUROC for discrimination, and predictive Brier where probability quality is directly evaluated. This appendix reports the additional threshold, ranking, and calibration diagnostics used to interpret those results.

\subsection{OLID and HateXplain Threshold Diagnostics}

Table~\ref{tab:appendix-binary-choice-diagnostics} expands the binary Choice results reported in the main text with positive-class precision and recall. These measures help characterize the operating point of each model but are not used as primary endpoints.

\begin{table}[p]
\centering
\scriptsize
\setlength{\tabcolsep}{1.2mm}
\caption{Detailed binary Choice offensiveness results on the cleaned OLID and HateXplain populations. Precision, recall, and F1 treat offensive content as the positive class. Detoxify predicts offensive when at least one category score is strictly greater than 0.5.}
\label{tab:appendix-binary-choice-diagnostics}
\resizebox{0.82\textwidth}{!}{%
\begin{tabular}{llrrrr}
\toprule
System & Information / variant & Acc. & Prec. & Rec. & F1 \\
\midrule
\multicolumn{6}{l}{\textbf{Panel A: OLID (14,046 posts)}} \\
\midrule
Jev & Direct & 74.4 & 58.7 & 75.8 & 66.1 \\
Jev & RAD-semantic & 73.5 & 58.2 & 69.3 & 63.3 \\
Laya & Direct & 72.6 & 62.2 & 43.0 & 50.8 \\
Laya & RAD-semantic & 59.6 & 43.8 & 80.1 & 56.6 \\
\midrule
Detoxify & Original & 79.7 & 76.3 & 55.6 & 64.3 \\
Detoxify & Original-small & 79.5 & 78.4 & 52.2 & 62.7 \\
Detoxify & Unbiased & 80.6 & 81.0 & 53.5 & 64.4 \\
Detoxify & Unbiased-small & 80.0 & 83.6 & 48.8 & 61.6 \\
Detoxify & Multilingual & 81.0 & 81.0 & 55.2 & 65.7 \\
\midrule
\multicolumn{6}{l}{\textbf{Panel B: HateXplain offensive/normal subset (14,123 posts)}} \\
\midrule
Jev & Direct & 59.8 & 51.2 & 90.7 & 65.5 \\
Jev & RAD-semantic & 66.1 & 57.1 & 78.0 & 65.9 \\
Laya & Direct & 59.6 & 72.3 & 6.1 & 11.3 \\
Laya & RAD-semantic & 56.0 & 48.4 & 71.4 & 57.7 \\
\midrule
Detoxify & Original & 58.2 & 50.2 & 65.9 & 57.0 \\
Detoxify & Original-small & 59.5 & 51.5 & 62.8 & 56.6 \\
Detoxify & Unbiased & 57.9 & 49.9 & 64.6 & 56.3 \\
Detoxify & Unbiased-small & 58.4 & 50.4 & 58.8 & 54.3 \\
Detoxify & Multilingual & 57.6 & 49.6 & 66.0 & 56.7 \\
\bottomrule
\end{tabular}%
}
\end{table}

The additional metrics clarify the operating-point differences summarized in the main text. On OLID, Jev Direct obtains substantially higher recall than the Detoxify variants, whereas Laya shifts toward higher recall under RAD at the cost of accuracy. The contrast is particularly pronounced on HateXplain: Laya Direct is highly conservative, with 72.3\% precision but only 6.1\% recall, while RAD moves it toward a substantially more permissive decision boundary.

\subsection{Noul Probability Diagnostics}

Table~\ref{tab:appendix-binary-noul-diagnostics} reports the complete diagnostics for the matched Direct--RAD Noul experiment. In addition to the main-text Accuracy, F1, and AUROC measures, we report precision, recall, average precision (AP), predictive Brier score, ten-bin Expected Calibration Error (ECE), and the percentage of posts classified as offensive (Pos.). AUROC and AP evaluate the continuous Noul probability, while the classification measures and Pos. use the fixed threshold $p(\mathrm{Offensive})>0.5$.

\begin{table}[p]
\centering
\scriptsize
\setlength{\tabcolsep}{0.65mm}
\caption{Detailed Direct and semantic-RAD Noul offensiveness results. Accuracy, precision, recall, F1, ECE, and Pos. are percentages; AUROC, AP, and predictive Brier are on $[0,1]$. Higher values are better except for Brier and ECE; Pos. is descriptive.}
\label{tab:appendix-binary-noul-diagnostics}
\resizebox{\textwidth}{!}{%
\begin{tabular}{llrrrrrrrrr}
\toprule
Model & Information & Acc. & Prec. & Rec. & F1 & AUROC & AP & Brier & ECE & Pos. \\
\midrule
\multicolumn{11}{l}{\textbf{Panel A: OLID (14,046 posts)}} \\
\midrule
Jev & Direct & 69.7 & 52.8 & 75.9 & 62.3 & .789 & .648 & .204 & 14.9 & 47.3 \\
Jev & RAD-semantic & 74.0 & 58.6 & 71.8 & 64.6 & .816 & .693 & .172 & 9.7 & 40.4 \\
Laya & Direct & 74.5 & 60.8 & 63.4 & 62.0 & .800 & .603 & .176 & 7.8 & 34.3 \\
Laya & RAD-semantic & 52.9 & 40.5 & 91.3 & 56.1 & .702 & .462 & .334 & 34.6 & 74.3 \\
\midrule
\multicolumn{11}{l}{\textbf{Panel B: HateXplain offensive/normal subset (14,123 posts)}} \\
\midrule
Jev & Direct & 56.3 & 49.0 & 94.4 & 64.5 & .749 & .651 & .326 & 33.2 & 81.1 \\
Jev & RAD-semantic & 63.3 & 54.0 & 86.0 & 66.3 & .755 & .667 & .245 & 20.2 & 67.0 \\
Laya & Direct & 56.1 & 48.5 & 70.1 & 57.3 & .604 & .481 & .304 & 23.0 & 60.8 \\
Laya & RAD-semantic & 43.3 & 42.5 & 98.4 & 59.3 & .515 & .409 & .494 & 49.5 & 97.4 \\
\bottomrule
\end{tabular}%
}
\end{table}

The probability diagnostics reinforce the distinction between discrimination and operating point. For Jev, retrieval improves AUROC, AP, predictive Brier, and ECE on OLID and improves the corresponding probability measures on HateXplain while reducing the fraction of posts classified as offensive. For Laya, RAD instead shifts the predicted class distribution strongly toward the positive class. The predicted-offensive rate rises from 34.3\% to 74.3\% on OLID and from 60.8\% to 97.4\% on HateXplain. This shift explains why recall can increase despite deteriorating AUROC, predictive Brier, and calibration.

\subsection{Detoxify Category-Vector Sensitivity}

Beyond binary classification, we explore whether SOM typed outputs reproduce the category structure of specialized toxicity detectors. Each model receives the post alone and independently evaluates the six categories shared by all Detoxify variants: toxicity, severe toxicity, obscenity, threat, insult, and identity attack. Noul returns a probability for each proposition, while Score places each category between absence and presence anchors. Neither interface receives Detoxify predictions, category definitions, written rules, or retrieved precedents.

Table~\ref{tab:detoxify-vector} compares the resulting six-dimensional vectors with Detoxify. Cosine similarity captures agreement in the relative category profile, while MAE captures disagreement in score magnitude. Jev aligns more closely with Detoxify than Laya on both OLID and HateXplain, although neither Noul nor Score dominates on both similarity measures.

\begin{table}[p]
\centering
\scriptsize
\setlength{\tabcolsep}{2mm}
\caption{Agreement between SOM category-score vectors and Detoxify outputs over the six heads shared by all five Detoxify variants. Cos. is mean per-post cosine similarity and MAE is mean per-post absolute error, averaged over the five reference variants. Higher cosine and lower MAE are better.}
\label{tab:detoxify-vector}
\resizebox{0.70\textwidth}{!}{%
\begin{tabular}{llrrrr}
\toprule
& & \multicolumn{2}{c}{OLID} & \multicolumn{2}{c}{HateXplain} \\
\cmidrule(lr){3-4}\cmidrule(l){5-6}
Model & Interface & Cos. & MAE & Cos. & MAE \\
\midrule
Jev & Noul & .696 & .263 & .728 & .353 \\
Jev & Score & .679 & .253 & .733 & .366 \\
Laya & Noul & .411 & .322 & .593 & .458 \\
Laya & Score & .565 & .554 & .630 & .476 \\
\bottomrule
\end{tabular}%
}
\end{table}

The main analysis compares SOM outputs with the six toxicity dimensions shared by all five Detoxify variants. As a sensitivity analysis, we repeat the comparison over all seven dimensions for the three Detoxify variants that additionally expose a sexually-explicit-content score: Unbiased, Unbiased-small, and Multilingual.

\begin{table}[p]
\centering
\scriptsize
\setlength{\tabcolsep}{2mm}
\caption{Seven-category sensitivity analysis of agreement between SOM outputs and Detoxify. Values are averaged over the Unbiased, Unbiased-small, and Multilingual Detoxify variants. Higher cosine similarity and lower MAE indicate stronger agreement.}
\label{tab:appendix-detoxify-seven}
\resizebox{0.70\textwidth}{!}{%
\begin{tabular}{llrrrr}
\toprule
& & \multicolumn{2}{c}{OLID} & \multicolumn{2}{c}{HateXplain} \\
\cmidrule(lr){3-4}\cmidrule(l){5-6}
Model & Interface & Cos. & MAE & Cos. & MAE \\
\midrule
Jev & Noul & .739 & .232 & .746 & .320 \\
Jev & Score & .724 & .223 & .750 & .333 \\
Laya & Noul & .407 & .338 & .568 & .481 \\
Laya & Score & .557 & .559 & .613 & .484 \\
\bottomrule
\end{tabular}%
}
\end{table}

The seven-category results preserve the ordering observed in the six-category analysis: Jev remains closer to the Detoxify reference vectors than Laya, while neither Noul nor Score dominates simultaneously in cosine similarity and absolute score magnitude.

\subsection{Published ModerationBench Reference Results}

The published ModerationBench systems are not included in the primary result table because they are not evaluated under the same input conditions as our SOM experiments. For completeness, Table~\ref{tab:appendix-moderationbench-published} reports the published stratum-level results used as external reference points.

\begin{table}[p]
\centering
\scriptsize
\setlength{\tabcolsep}{1.6mm}
\caption{Published ModerationBench reference results on the original multimodal benchmark. Random, Moderated, and Near-moderated report positive-class F1; Safe reports accuracy because the stratum contains no positive targets. These results are descriptive external references and are not directly comparable with our text-plus-caption adaptation.}
\label{tab:appendix-moderationbench-published}
\resizebox{0.78\textwidth}{!}{%
\begin{tabular}{llrrrr}
\toprule
System & Context & Random F1 & Moderated F1 & Near-moderated F1 & Safe Acc. \\
\midrule
gemma4 & Full policy & 52.0 & 92.0 & 81.0 & 100.0 \\
qwen3.5 & Full policy & 52.0 & 92.0 & 77.0 & 100.0 \\
gemini3.5 & Full policy & 60.0 & 92.0 & 83.0 & 100.0 \\
gpt5.6 & Full policy & 41.0 & 92.0 & 79.0 & 100.0 \\
\midrule
gemma4 & Prototype & 47.0 & 91.0 & 82.0 & 100.0 \\
gemini3.5 & Prototype & 59.0 & 93.0 & 84.0 & 100.0 \\
gpt5.6 & Prototype & 42.0 & 91.0 & 76.0 & 99.1 \\
BMS & Deployed & 22.0 & 91.0 & 0.0 & 100.0 \\
\bottomrule
\end{tabular}%
}
\end{table}

The published systems process the original media and receive either explicit policy information or separately constructed multimodal examples, whereas our SOM evaluation uses text combined with generated neutral captions and five leave-one-out binary precedents. The table should therefore be interpreted as contextual evidence about benchmark difficulty rather than as an input-matched model comparison.

\section{ModerationBench Caption-Model Sensitivity}
\label{app:moderationbench-captions}

The main ModerationBench experiment uses neutral media descriptions generated by \texttt{OpenGVLab/InternVL3\_5-4B-Instruct}. The source snapshot also provides independently generated descriptions from \texttt{Qwen/Qwen2.5-VL-3B-Instruct} and \texttt{openbmb/MiniCPM-V-4\_5}. We repeat all four Direct/RAD and Choice/Noul conditions for both SOMs with each caption representation. As predeclared for this extension, each caption model uses its maximum deterministic clean population rather than a shared intersection. The resulting populations contain 3,673 Qwen posts (2,580 Safe and 1,093 Unsafe), 3,689 InternVL posts (2,584 Safe and 1,105 Unsafe), and 3,685 MiniCPM posts (2,583 Safe and 1,102 Unsafe). The differences arise because caption availability and the exact text used for duplicate and contradiction detection vary by caption model. Consequently, comparisons across caption models combine a representation effect with a small population-composition effect; Direct--RAD comparisons within one caption population remain exactly matched.

Table~\ref{tab:appendix-caption-pooled} reports the complete pooled results. MiniCPM yields the strongest Jev accuracy and F1 in every corresponding condition. Its RAD-Noul configuration reaches 89.1\% accuracy, 82.5\% F1, .950 AUROC, and .890 average precision. The retrieval effect is nevertheless consistent across caption generators: Jev RAD raises Choice accuracy by 2.8, 2.7, and 1.6 percentage points for Qwen, InternVL, and MiniCPM, respectively, and raises Noul accuracy by 3.2, 3.0, and 2.2 points. Laya also behaves consistently across caption generators. RAD-Choice reaches approximately 66\% accuracy in all three populations, whereas RAD-Noul reaches 95.8--96.3\% recall but only 50.5--52.0\% accuracy because it predicts Unsafe too often.

\begin{table}[p]
\centering
\scriptsize
\setlength{\tabcolsep}{0.5mm}
\caption{Pooled ModerationBench results under all three neutral-caption generators. D and R denote Direct and semantic RAD; C and N denote Choice and Noul. Accuracy, precision, recall, F1, and ECE are percentages; AUROC, AP, and predictive Brier are on $[0,1]$. Higher is better except for Brier and ECE. Populations are caption-specific and therefore not exactly matched across generators.}
\label{tab:appendix-caption-pooled}
\resizebox{\textwidth}{!}{%
\begin{tabular}{lllrrrrrrrrr}
\toprule
Caption & SOM & Cond. & $n$ & Acc. & Prec. & Rec. & F1 & AUROC & AP & Brier & ECE \\
\midrule
Qwen & Jev & D-C & 3,673 & 84.1 & 73.7 & 72.6 & 73.1 & .893 & .783 & .121 & 6.8 \\
Qwen & Jev & R-C & 3,673 & 86.9 & 75.7 & 82.4 & 78.9 & .933 & .852 & .097 & 5.5 \\
Qwen & Jev & D-N & 3,673 & 84.6 & 78.2 & 67.1 & 72.2 & .900 & .800 & .112 & 2.4 \\
Qwen & Jev & R-N & 3,673 & 87.9 & 78.7 & 81.2 & 79.9 & .936 & .861 & .095 & 7.1 \\
\addlinespace
Qwen & Laya & D-C & 3,673 & 47.2 & 30.7 & 61.8 & 41.0 & .524 & .313 & .334 & 30.2 \\
Qwen & Laya & R-C & 3,673 & 66.3 & 44.0 & 48.9 & 46.3 & .663 & .382 & .252 & 15.4 \\
Qwen & Laya & D-N & 3,673 & 63.1 & 41.0 & 54.7 & 46.9 & .651 & .424 & .312 & 29.8 \\
Qwen & Laya & R-N & 3,673 & 51.6 & 37.7 & 96.2 & 54.2 & .721 & .444 & .437 & 45.5 \\
\midrule
InternVL & Jev & D-C & 3,689 & 84.0 & 73.5 & 72.9 & 73.2 & .893 & .785 & .121 & 6.8 \\
InternVL & Jev & R-C & 3,689 & 86.7 & 74.8 & 84.0 & 79.1 & .936 & .860 & .097 & 5.4 \\
InternVL & Jev & D-N & 3,689 & 84.4 & 78.4 & 66.2 & 71.8 & .902 & .804 & .112 & 3.0 \\
InternVL & Jev & R-N & 3,689 & 87.4 & 77.6 & 81.7 & 79.6 & .940 & .867 & .094 & 7.1 \\
\addlinespace
InternVL & Laya & D-C & 3,689 & 46.5 & 31.3 & 65.8 & 42.4 & .521 & .308 & .344 & 30.9 \\
InternVL & Laya & R-C & 3,689 & 66.0 & 43.9 & 49.0 & 46.3 & .661 & .385 & .255 & 16.3 \\
InternVL & Laya & D-N & 3,689 & 62.2 & 40.4 & 55.2 & 46.7 & .658 & .427 & .313 & 29.6 \\
InternVL & Laya & R-N & 3,689 & 50.5 & 37.3 & 96.3 & 53.8 & .723 & .454 & .447 & 46.4 \\
\midrule
MiniCPM & Jev & D-C & 3,685 & 86.8 & 75.8 & 82.1 & 78.8 & .930 & .839 & .099 & 4.8 \\
MiniCPM & Jev & R-C & 3,685 & 88.4 & 77.1 & 86.9 & 81.7 & .947 & .883 & .088 & 5.6 \\
MiniCPM & Jev & D-N & 3,685 & 86.9 & 79.6 & 75.6 & 77.5 & .932 & .848 & .095 & 4.8 \\
MiniCPM & Jev & R-N & 3,685 & 89.1 & 79.0 & 86.5 & 82.5 & .950 & .890 & .086 & 7.8 \\
\addlinespace
MiniCPM & Laya & D-C & 3,685 & 46.8 & 31.0 & 63.4 & 41.6 & .518 & .311 & .341 & 30.9 \\
MiniCPM & Laya & R-C & 3,685 & 66.4 & 44.5 & 49.1 & 46.7 & .670 & .390 & .248 & 15.3 \\
MiniCPM & Laya & D-N & 3,685 & 63.3 & 41.6 & 56.4 & 47.9 & .663 & .435 & .309 & 29.1 \\
MiniCPM & Laya & R-N & 3,685 & 52.0 & 38.0 & 95.8 & 54.4 & .735 & .467 & .434 & 45.1 \\
\bottomrule
\end{tabular}%
}
\end{table}

\bibliographystyle{abbrv}
\bibliography{bib_ida}

\end{document}